\documentclass[runningheads]{llncs}
\usepackage[nocompress]{cite}
\usepackage{float}
\usepackage{graphicx}
\usepackage{booktabs}
\usepackage{multirow}
\usepackage[caption=false,font=footnotesize,labelfont=sf,textfont=sf]{subfig}
\DeclareGraphicsExtensions{.png,.jpg}
\usepackage{physics}
\usepackage{enumitem}
\setlist[itemize]{leftmargin=*}

\begin{document}
\title{A Survey on Map-Matching Algorithms}

\author{Pingfu Chao\inst{1} \and
Yehong Xu\inst{1} \and
Wen Hua\inst{1} \and
Xiaofang Zhou\inst{1}}
\authorrunning{P. Chao et al.}
%
\institute{School of Information Technology and Electrical Engineering, \\
	The University of Queensland, Australia \\
\email{\{p.chao,yehong.xu,w.hua\}@uq.edu.au, zxf@itee.uq.edu.au}}
\maketitle
\vspace{-1em}
\begin{abstract}
	The map-matching is an essential preprocessing step for most of the trajectory-based applications. Although it has been an active topic for more than two decades and, driven by the emerging applications, is still under development. There is a lack of categorisation of existing solutions recently and analysis for future research directions. In this paper, we review the current status of the map-matching problem and survey the existing algorithms. We propose a new categorisation of the solutions according to their map-matching models and working scenarios. In addition, we experimentally compare three representative methods from different categories to reveal how matching model affects the performance. Besides, the experiments are conducted on multiple real datasets with different settings to demonstrate the influence of other factors in map-matching problem, like the trajectory quality, data compression and matching latency.
\end{abstract}

\vspace{-2em}
\section{Introduction}\label{sec:intro}

Nowadays, the ubiquity of positioning devices enables the tracking of user/vehicle trajectories. However, due to the intrinsic inaccuracy of the positioning systems, a series of preprocessing steps are required to correct the trajectory errors. As one of the major preprocessing techniques, the map-matching algorithm finds the object’s travel route by aligning its positioning data to the underlying road network. It is the prerequisite of various location-based applications, such as navigation, vehicle tracking, map update and traffic surveillance.

The map-matching problem has been studied for more than two decades. Despite hundreds of papers are proposed, to the best of our knowledge, only several works were conducted\cite{quddus2007current,hashemi2014critical,wei2013mapc,kubicka2018comparative} surveying them. More importantly, even the most recent surveys\cite{kubicka2018comparative} fail to categorise the existing methods comprehensively. They either classify them based on applications\cite{kubicka2018comparative} that are not very distinctive to each other, or follow the previous categorisation\cite{quddus2007current} that is obsolete. Besides, various new techniques are introduced to the map-matching problem recently, including new models (weight-based\cite{sharath2019dynamic}, multiple hypothesis theory\cite{taguchi2018online}), new tuning techniques (machine learning\cite{osogami2013map}, information fusion\cite{li2013high,hu2017if}), new data types (DGPS, inertial sensor, semantic road network) and new research topics (lane-level, parallel). Hence, it is about time to conduct a new survey to summarise existing solutions and provide guidance to future research.

Note that the existing map-matching problem covers various scenarios, ranging from indoor to outdoor and from pedestrian, vehicle to multimodal. However, to ensure a unified setting for survey and comparison, in this paper, we target the vehicle trajectory map-matching in an outdoor environment due to its popularity. We categorise the existing work from technical perspective. In addition, we discuss the main properties of the methods and future research directions according to the experiment results conducted on multiple matching algorithms. Overall, our contributions are listed as follows:

\begin{itemize}
	\item We review the map-matching solutions proposed since the last comprehensive survey\cite{quddus2007current} and propose a new categorisation of the algorithms based on their methodology. Our proposed categorisation can better distinguish the existing methods from the technical perspective, which is beneficial for future study.
	
	\item We enumerate several map-matching challenges that are caused by low-quality trajectory data. The challenges are exemplified and explained concretely, which leads to future research directions.
	
	\item To further demonstrate the challenges, we implement three representative map-matching algorithms and conduct extensive experiments on datasets with different sampling rate, map density and compression level. Our claims about the relationship between data quality and map-matching quality are fully supported by the experiments.
\end{itemize}

The rest of the paper is organised as follows: In Section \ref{sec:preliminaries}, we first formally define the map-matching problem and enumerate the existing surveys and their limitations. Then, we propose our new categorisation in Section \ref{sec:survey}. We further discuss the current challenges which are demonstrated through experiments in Section \ref{sec:experiment} and we draw conclusions in Section \ref{sec:conclusion}.

\vspace{-1em}
\section{Preliminaries}\label{sec:preliminaries}

\subsection{Problem Definition}

We first define the map-matching problem and relevant datasets, including trajectory (input), road network (input) and route (output):

\begin{definition}(Trajectory)\label{def:trajectory}
	A \textbf{trajectory} $Tr$ is a sequence of chronologically ordered spatial points $Tr:p_{1} \rightarrow p_{2} \rightarrow ... \rightarrow p_{n}$ sampled from a continuously moving object. Each point $p_{i}$ consists of a 2-dimensional coordinate $<x_{i},y_{i}>$, a timestamp $t_{i}$, a speed $spd_{i}$ (optional) and a heading $\theta_{i}$ (optional). i.e.: $p_{i} = <x_{i},y_{i},t_{i},spd_{i},\theta_{i}>$.
\end{definition}

\begin{definition}(Road Network)
	A \textbf{road network} (also known as map) is a directed graph $G = (V,E)$, in which a vertex $v = (x,y) \in V$ represents an intersection or a road end, and an edge $e = (s,e,l)$ is a directed road starting from vertices $s$ to $e$ with a polyline $l$ represented by a sequence of spatial points.
\end{definition}

\begin{definition}(Route)
	A route $R$ represents a sequence of connected edges, i.e. $R: e_{1} \rightarrow e_{2} \rightarrow ... \rightarrow e_{n}$, where $e_{i} \in G.E (1 \leq i \leq n)$ and $e_{k}.e = e_{k+1}.s$.
\end{definition}

\begin{definition}(Map-Matching)\label{def:map-matching}
	Given a road network $G(V,E)$ and a trajectory $Tr$, the map-matching find a route $\mathcal{MR}_{G}(Tr)$ that represents the sequence of roads travelled by the trajectory.
\end{definition}

For simplicity, we omit the subscript $G$ and use $\mathcal{MR}(Tr)$ instead to represent the matching result as different trajectories are usually map-matched on the same map. In general, the map-matching route is expected to be continuous as it represents the vehicle's travel history. However, it is quite often that $\mathcal{MR}(Tr)$ contains disconnected edges due to incorrect map-matching, which will be discussed in Section \ref{sec:experiment}.

\subsection{Related Work}

Intuitively, since the vehicle usually runs on the roads, a fully accurate trajectory sampled from a vehicle should always lie on the map. Therefore, apart from some unexpected map errors, which happens less frequently and is addressed by map update process\cite{chao2019trajectories}, the difficulty of map-matching problem solely depends on the quality of the input trajectories. As studied in many papers, the quality issues in trajectories are pervasive, which mainly caused by inaccurate measurement and low sampling rate. In terms of the \textit{measurement error}, due to the unstable connection between GPS device and satellites, the location of GPS samples usually deviate from its actual position by a random distance. Meanwhile, the \textit{sampling error} is mainly caused by lowering the sampling frequency.



To deal with the quality issues, the map-matching problem has been studied for more than two decades. In terms of the working scenarios and applications, the current map-matching solutions can be classified into online mode and offline mode. In online map-matching, the vehicle positions are sampled continuously and are processed in a streaming fashion, which means each time the map-matching is only performed on the current sample with a limited number of preceding or succeeding samples\cite{goh2012online,yin2018feature} as reference. The process is usually simple and fast for interactive performance. In contrary, the offline map-matching is performed after the entire trajectory is obtained, so it aims for optimal matching route with less constraint on processing time.


From the methodology perspective, Quddus et al.\cite{quddus2007current} first conducted a comprehensive review of the map-matching algorithms proposed before 2007. The paper classified the methods into four categories, namely \textit{geometric}, \textit{topology}, \textit{probabilistic} and \textit{advanced}. The \textit{geometric} methods only focus on the distance between trajectory elements and the road network, while the \textit{topology} methods take into consideration the connectivity and shape similarity. The \textit{probabilistic} methods try to model the uncertainty of trajectory, including the measurement error and the unknown travel path between two samples, and they aim to find a path that has the highest probability to generate the given trajectory. The \textit{advanced} category contains methods that are based on some advanced models, like Kalman Filter, particle filter and fuzzy logic. This categorisation shows the evolution of map-matching research, which starts from simple, fast but inaccurate geometric-based methods to more complicated but accurate probability/advanced solutions. It is by far the most comprehensive survey of this field. However, after more than ten years' development, most of the methods mentioned in the paper has been outperformed by their new successors and the previous categorisation also requires a revisit. Several surveys proposed afterwards reviewed the methods in certain perspectives. Hashemi et al. \cite{hashemi2014critical} targeted at the online map-matching scenario. Kubi\v{c}ka et al. discussed the map-matching problem based on the applications\cite{kubicka2018comparative}, namely \textit{navigation}, \textit{tracking} and \textit{mapping}. Other categorisations also appear recently (\textit{incremental max-weight}, \textit{global max-weight} and \textit{global geometry}\cite{wei2013mapc}) which shows that there is still no consensus on how to classify the algorithms technically. However, all of the existing categorisations inherit the same idea from Quddus' survey\cite{quddus2007current} with minor variations, which fail to categorise the recent methods for multiple reasons, explained in Section \ref{sec:survey}.

\vspace{-1em}

\section{Survey of Map-Matching Algorithm}\label{sec:survey}

According to our study, previous categorizations fail to classify the current solutions due to three main reasons: (1) Categories for some primary methods, such as \textit{geometric} category\cite{quddus2007current}, are no longer the focus due to their weak performance. (2) Application-based classification\cite{hashemi2014critical,kubicka2018comparative} cannot fully distinguish the methods. Many of the map-matching algorithms, like the Hidden Markov Model (HMM) and Multiple Hypothesis Technique (MHT), apply to both online and offline scenarios for different applications. (3) Classifying algorithms by embedded mathematical tools are not feasible since many recent algorithms employ multiple mathematical tools. Furthermore, the same tool implemented in different algorithms may be used for different purposes, for example, an extended Kalman filter can be used to either estimate biases in GPS or fuse measurements from different sources\cite{li2013high}. 

Therefore, we establish a new classification that classifies the map-matching algorithms by their core matching model, which is employed to coordinate their techniques to finally achieve map-matching. In a map-matching algorithm, the map-matching model is the overall framework or matching principle for the map-matching process. A model usually consists of a set of computation components, like the calculation of distance, transition and user behaviour modelling, and a workflow connecting them. Those components are fixed while their definition and implementation vary among different methods. Existing map-matching models can be categorised into four classes: \textit{similarity model}, \textit{state-transition model}, \textit{candidate-evolving model} and \textit{scoring model}.

\subsection{Similarity Model}

The similarity model refers to a general approach that returns the vertices/edges that is \textit{closest} to the trajectory geometrically and/or topologically. Intuitively, since a vehicle's movement always follows the topology of the underlying road network and the vehicle can never leap from one segment to another, the trajectory should also similar to those of the true path on the map. Therefore, the main focus in this category is how to define the \textit{closeness}.

\subsubsection{Distance-based}

Most of the earliest point-to-curve and curve-to-curve matching algorithms\cite{quddus2007current} follow this idea. Specifically, the point-to-curve solution projects each trajectory point to the geometric-closest edge, whereas the curve-to-curve matching algorithms project each trajectory segment to the closest edge where the \textit{closeness} is defined by various similarity metrics. Fr\'echet distance is the most commonly-used distance function\cite{wei2013mapf} since it considers the monotonicity and continuity of the curves. However, it is sensitive to trajectory measurement errors since its value can be dominated by the outliers. As an alternative, Longest Common Subsequence (LCSS)\cite{zhu2017trajectory} divide a trajectory into multiple segments and find the shortest path on the map for each pair of start and end points of a trajectory segment. The shortest paths are then concatenated and form the final path while their corresponding LCSS scores are summed as the final score. Then, the path whose LCSS score is higher than a predefined threshold is regarded as the final matching result.

\subsubsection{Pattern-based}

The pattern-based algorithms utilise the historical map-matched data to answer new map-matching queries by finding similar travel patterns\cite{zheng2012reducing}. The assumption is that people tend to travel on the same paths given a pair of origin and destination points. Therefore, by referring to the historical trajectories that are similar to the query trajectory, its candidate paths can be obtained without worrying about the sparseness of trajectory samples. Specifically, a historical trajectory or a trajectory obtained by concatenating multiple historical trajectories will be referred to if each point of this trajectory is in the safe region around the query trajectory. The algorithm finally uses a scoring function to decide the optimal route. However, due to the sparsity and disparity of historical data, the query trajectory may not be fully covered by historical trajectories especially in some rarely travelled regions, which leads to a direct matching process.


\subsection{State-Transition Model}

The state-transition models build a weighted topological graph which contains all possible routes the vehicle might travel. In this graph, the vertices represent the possible \textit{states} the vehicle may be located at a particular moment, while the edges represent the \textit{transitions} between states at different timestamps. Different from the road network, the weight of a graph element represents the possibility of a state or a transition, and the best matching results comes from the optimal path in the graph globally. There are three major ways of building the graph and solving the optimal path problem, namely Hidden Markov model (HMM), Conditional Random Field (CRF) and the Weighted Graph Technique (WGT).

\subsubsection{Hidden Markov model}

HMM is one of the most widely used map-matching models as it simulates the road network topology meanwhile considers the reasonability of a path. HMM focuses on the case when states in the Markov chain are unobservable (hidden) but can be estimated according to the given observations associated with them. This model fits in the map-matching process naturally. Each trajectory sample is regarded as the observation, while the vehicle actual location on the road, which is unknown, is the hidden states. In fact, due to the trajectory measurement error, all the roads near the observation can potentially be the actual vehicle location (state), each of which with a probability (emission probability). As the trajectory travels continuously, the transition between two consecutive timestamps is concluded by the travel possibility (transition probability) between their candidate states. Therefore, the objective is to find an optimal path which connects one candidate in every timestamp. The final path is obtained by the Viterbi algorithm which utilises the idea of dynamic programming. The major difference between various HMM-based algorithms is their definition of emission probability and transition probability. Unlike the emission probability, which is defined identically in most papers, the definition of the transition probability varies since the travel preference can be affected by plenty of factors. Some works\cite{newson2009hidden} prefers a candidate pair whose distance is similar to the distance between the observation pair, while others consider velocity changes\cite{goh2012online}, turn restriction\cite{osogami2013map}, closeness to the shortest path, the heading mismatch and travel penalty on U-turns, tunnels and bridges. Besides, HMM is also applied to online scenario\cite{goh2012online}. However, to build a reasonable Markov chain, online HMM-based algorithms usually suffer from latency problems, which means a point is matched after a certain delay.

\subsubsection{Conditional random field}

CRF is utilized in many areas as an alternative to HMM to avoid the selection bias problem\cite{hunter2014path}. As both CRF and HMM are statistic models, the major difference is that CRF models interactions among observations while HMM models only model the relation between an observation with the state at the same stage and its closest predecessor. Hunter et al.\cite{hunter2014path} proposed a CRF-based map-matching algorithm that can be applied to both online and offline situations with high accuracy. Its overall approach is similar to HMM-based algorithms but with different transition probability which considers the maximum speed limit and the driving patterns of drivers. However, the problem shared by both HMM and CRF is the lack of a recovery strategy for the match deviation. Since once a path is confirmed, it will be contained by all future candidate paths, which hurts the online scenarios especially.

\subsubsection{Weighted graph technique}

WGT refers to a model that infers the matching path from a weighted candidate graph, where the nodes are candidate road points of location measurements and edges are only formed between two nodes corresponding to two consecutive samples. In most WGT-based algorithms, candidate points are the closest points on road segments in a radius of measurements\cite{lou2009map,hu2017if}, which is similar to HMM. The process of the WGT can be summarized as three steps: (1) Initializing the candidate graph. (2) Weighting edges in the graph using a scoring function. (3) Inferring a path based on the weighted graph.

Algorithms fall in this category mainly differs from each other in weighting functions. Lou et al.\cite{lou2009map} firstly propose the WGT. It weights an edge simply based on a spatial cost and a temporal cost, where the spatial cost is modelled on the distance between candidate $c_{i}$ to its observed position $p_{i}$ and the shortest length between $c_{i}$ and $c_{i+1}$ whereas the temporal cost is modelled on the velocity reasonability. Based on Lou's design, the following work further considers mutual influences between neighbouring nodes, road connectivity, travel time reasonability\cite{hu2017if} and other road features (traffic lights, left turns, etc.).


\subsection{Candidate-Evolving Model}

Candidate-evolving model refers to a model which holds a set of candidates (also known as particles or hypotheses) during map-matching. The candidate set is initiated based on the first trajectory sample and keeps evolving by adding new candidates propagated from old ones close to the latest measurements while pruning irrelevant ones. Interpreting a candidate as a vote, by maintaining the candidate set, the algorithms are able to find a segment with the most votes, thereby, determining the matching path. Compare to the state-transition model, the candidate-evolving model is more robust to the off-track matching issue since the current matching is influenced not only by a previously defined solution, but also by other candidates. The particle filter (PF) and the Multiple Hypothesis Technique (MHT) are two representative solutions. 

\subsubsection{Particle filter}

PF is a state estimation technique that combines Monte Carlo sampling methods with Bayesian Inference. This technique has been utilized to support map-matching by the way of sensor fusion and measurement correction\cite{wang2016improved}, while it is also applicable to directly address map-matching problem\cite{bonnifait2009multi}. The general idea of the PF model is to recursively estimate the Probability Density Function (PDF) of the road network section around the observation as time advances. Here, the PDF is approximated by $N$ discrete particles, each particle maintains a weight representing how consistent it is to the location observation. The process of a PF can be summarized as follows: Initially, $N$ particles are sampled with the same weight representing different locations in the local road network. The weight of each particle keeps getting updated as new observations are received. Then the PDF for the road network section around the new observations is calculated and the area with the highest probability is determined as the matched region. A resampling stage starts afterwards, where a new set of particles are derived based on the current set. The particles with higher weights are more likely to propagate according to moving status to feed particles for the next cycle, while those with low weights are likely to die out.


\subsubsection{Multiple hypothesis technique}

Similar to PF, the MHT also tries to maintain a list of candidate road matches for the initial trajectory point and the list is expected to be as large as possible to ensure correct result coverage. However, different from the PF which iterate through all possibilities over time, the MHT is a much simpler model that inherits the idea of maintaining hypotheses but manages to reduce computation during the process. An MHT evaluates each candidate road edge (or point) based on a scoring function instead of trying to approximate the complicated PDF for the neighbour map area. Thereby, the computation cost of the MHT is dramatically reduced. According to the intuition, the MHT can be easily adopted in online scenario\cite{taguchi2018online}. Moreover, since it possesses all the possibility of previous hypotheses, Taguchi et al.\cite{taguchi2018online} propose a prediction model which extends the hypotheses to further predict the future route, which can achieve better online map-matching accuracy without introducing latency.

\subsection{Scoring Model}

\subsubsection{Na\"{i}ve weighting}

A group of algorithms\cite{quddus2015shortest,sharath2019dynamic} apply the weight without using a particular model. Instead, they simply assign a group of candidates to each trajectory segment (or location observation) and find a road edge from each group that maximizes the predefined scoring function. The found segment in every timestamp is either returned if applied to the online scenario or waited to be joint with other matched segments if applied in the offline scenario. Most recent work in this category\cite{sharath2019dynamic} achieves a lane-level map-matching performance. The algorithm first identifies lanes in each road by utilising the road width information in the map and partition them into grids accordingly. The algorithm then finds candidate lane grids around the observed location and scores these grids at each timestamp. The grid results in the maximum score are then returned. The scoring function is a linear combination of four features, i.e. the proximity between the grid and trajectory sample, the estimated location of the vehicle at the next time stage, the reachability from the grid and the intention of a turn. These features are modelled individually, their scores can be obtained from the corresponding models in every timestamp. In addition, feature scores are weighted differently in the scoring function whose coefficients are computed by a training process before map-matching starts.

\vspace{-1em}
\section{Challenges and Evaluations}\label{sec:experiment}

Despite various of map-matching models are proposed to deal with trajectory quality issues, the current solutions still fail to achieve decent matching quality in all scenarios. Therefore, in this section, we will discuss several major challenges caused by data quality issues that are affecting the map-matching results. We will demonstrate them both visually and experimentally to exemplify their significance.

\subsection{Experimental Settings}

As listed in Table \ref{table:dataset summary}, we use four datasets for our experiments. The \textit{Global}\cite{kubivcka2015dataset} dataset is a public dataset for map-matching evaluation. It contains 100 GPS trajectories sampled from 100 different areas all over the world, each of which is provided with a dedicate underlying map. Besides, we extract three sub-areas, namely \textit{Beijing-U}, \textit{Beijing-R} and \textit{Beijing-M}, from a commercial dataset which contains taxi trajectories in Beijing. The reason for choosing these four datasets is their diversity in terms of trajectory quality and map density. The \textit{Global} dataset has the best trajectory accuracy and its maps are also very sparse. The \textit{Beijing-U} and \textit{Beijing-R} represent two maps extracted from urban and rural areas, respectively. They have roughly the same size but different map density ($27.3 vs 13.9$), so they can be used to evaluate the influence of map density to map-matching results. \textit{Beijing-M} is a larger map area with more trajectories for large-scale performance test.

\vspace{-1em}
\begin{table*}[htbp]
	\caption{Summary of experiment datasets}
	\centering
	\label{table:dataset summary}
	\scriptsize
	\begin{tabular}{cccccccc}
		\hline
		Name & \multicolumn{3}{c}{Input Trajectory} & \multicolumn{4}{c}{Road Network} \\
		\hline
		& \textit{Trajectory} & \textit{Trajectory} & \textit{Sampling} & \textit{\# of vertices} & \textit{\# of} & \textit{Map} & \textit{Map Density} \\
		& \textit{Count}  & \textit{Point Count} &\textit{rate(sec)}  & \textit{+ mini nodes} & \textit{edges} & \textit{Size($km^{2}$)} & ($km/km^{2}$) \\
		\hline
		Global & 100 & & 1 & N/A & N/A & N/A & N/A \\
		Beijing-U & 7,905 & 247,544 & 11.0 & 7,672 & 4,484 & 9.9 & 27.3 \\
		Beijing-R & 3,106 & 119,612 & 8.6  & 3,927 & 1,326 & 9.9 & 13.9 \\
		Beijing-M & 73,072 & 3,285,934 & 10.3 & 41,353 & 22,580 & 57.0 & 24.2 \\
	\end{tabular}
	\vspace{-2em}
\end{table*}

Our experiments are performed on a single server with two Intel(R) Xeon(R) CPU E5-2630 with 10 cores/20 threads at 2.2GHz each, 378GB memory and Ubuntu 16.04. Both the route matching result $\mathcal{MR}(Tr)$ and the corresponding ground-truth are regarded as sets of road edges and are evaluated by F-measure, which is commonly used in map-matching evaluation\cite{wei2013mapc,sharath2019dynamic}. The candidate map-matching algorithms used in the experiments include the most popular offline HMM map-matching\cite{newson2009hidden}, the most-recent offline WGT algorithm\cite{yang2018fast} and an online Scoring method\cite{quddus2015shortest}.

\subsection{Data Quality Challenges}

According to our observations from the experiments, the current data quality issues affect the map-matching in three major ways: the unnecessary detours, the matching breaks and the matching uncertainty. 

\subsubsection{Unnecessary detour}

\begin{figure*}[htbp]
	\centering
	\subfloat[Unnecessary detour]{\includegraphics[width=0.45\textwidth]{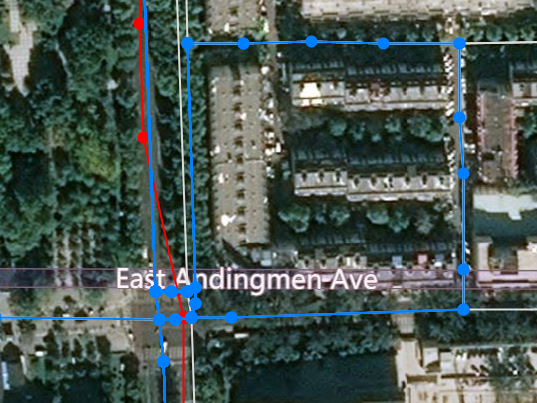}\label{fig:unecessary detour}}
	\subfloat[Matching break]{\includegraphics[width=0.45\textwidth]{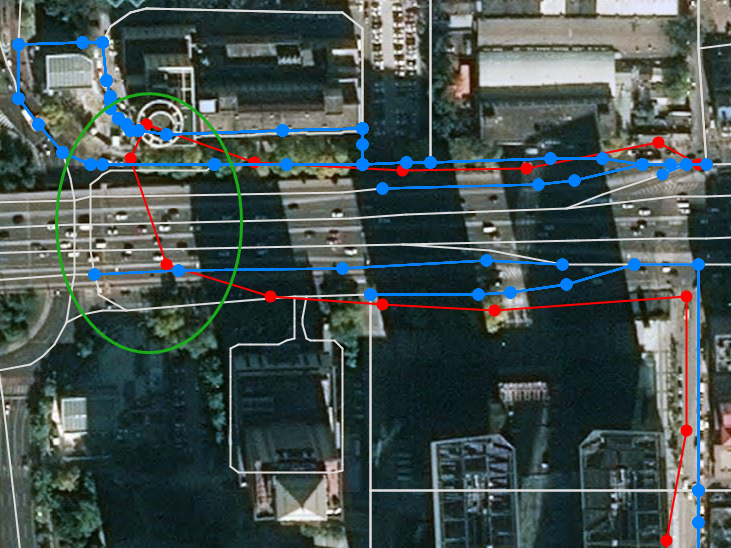}\label{fig:matching break}}
	\caption{Example of map-matching challenges}
\end{figure*}

As an example shown in Fig. \ref{fig:unecessary detour}, the matching result sometimes may contain unnecessary detours, which happens more frequently when the trajectory sampling rate is very high. In most scenarios, the detour is caused by two consecutive trajectory samples being too close to each other so that the succeeding point happens to be matched to the upper stream of its preceding point. Therefore, the shortest path between these two points has to go through a long detour. To avoid such issue, the measurement error should be considered when finding the shortest path, which means a certain degree of backtrace should be allowable. Alternatively, instead of simply project trajectory samples to the candidate roads to find candidate points, the actual matching point should follow a distribution, according to the trajectory measurement error, along the candidate road.  

\begin{figure*}[htbp]
	\centering
	\subfloat[Accuracy over different sampling rate]{\includegraphics[width=0.33\textwidth]{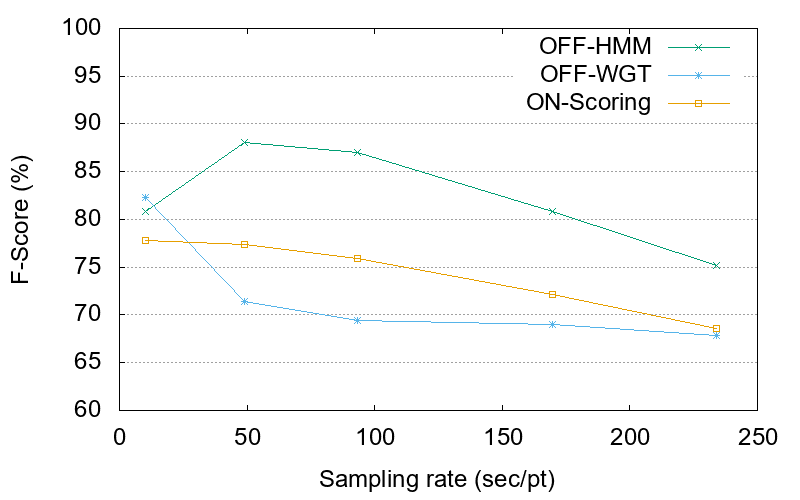}\label{fig:accuracy sampling}}
	\subfloat[Down-sample v.s. compression]{\includegraphics[width=0.33\textwidth]{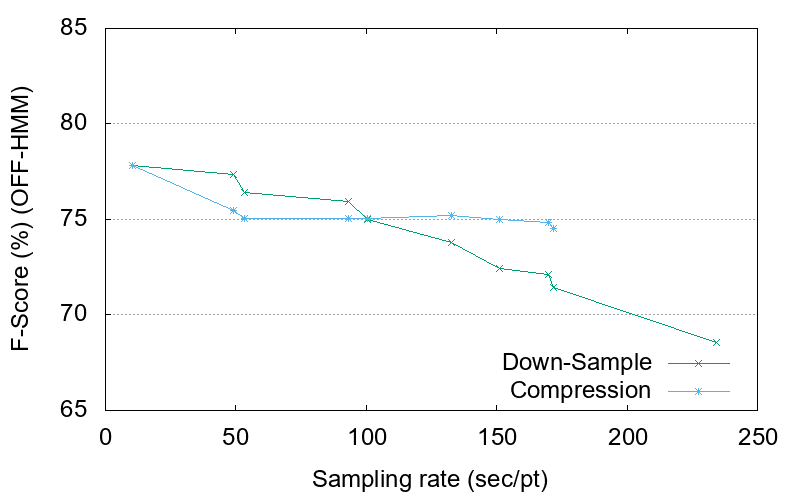}\label{fig:accuracy compression}}
	\subfloat[Influence of map density and trajectory quality]{\includegraphics[width=0.33\textwidth]{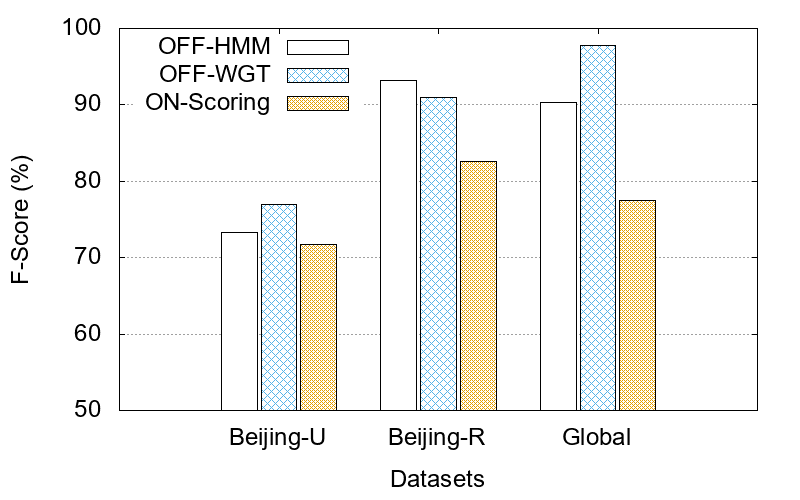}\label{fig:accuracy density}}
	\caption{Experimental results}
	\vspace{-2em}
\end{figure*}

In general, as depicted in Fig. \ref{fig:accuracy sampling}, the detour problem strongly affects the matching quality when the sampling rate is high. The result shows that it is not always the case that a higher sampling rate leads to higher matching quality especially when the measurement error becomes the major problem. Therefore, a better way of modelling the measurement error is still required. 

\vspace{-2em}

\vspace{-2em}
\subsubsection{Matching break}

The matching break is a common problem in map-matching, which is mainly caused by trajectory outliers. This happens more frequently in the state-transition matching model when the correct state falls out of the candidate range of the outlier. In this case, the states of two consecutive observations may be unreachable, leading to disconnected matching route, as shown in the green circled area in Fig. \ref{fig:matching break}. Currently, most of the solutions\cite{newson2009hidden} try to overcome this problem by identifying and removing the outliers to remedy the broken route. In Fig. \ref{fig:accuracy compression}, we apply online scoring method on \textit{Beijing-M} with random down-sample and trajectory compression (Douglas-Peucker algorithm), respectively. The result shows that simple trajectory compression fails to prune outliers as they are usually preserved as outstanding point, which means more preprocessing step is required to remove such outliers. However, considering the detour problem in high sampling rate data, the trajectory compression achieves better performance compared with simply down-sample the trajectory as it better preserve the shape of the trajectory, which is still beneficial. 

\subsubsection{Matching uncertainty}

Although the main goal of map-matching algorithms is to reduce the uncertainty of trajectory, the matching uncertainty varies in different scenarios. One of the main factor, which is not mentioned by any of previous work, is the map density. Intuitively, the map-matching of trajectory is much harder when the map area is full of roads compared with an emptier region. As shown in Fig. \ref{fig:accuracy density}, the map density can significantly affect the matching quality as the \textit{Beijing-U} has much worse performance than \textit{Beijing-R} given both of them have a similar trajectory quality. On the other hand, the trajectory quality also plays an important role since the performance on \textit{Global} is better than on \textit{Beijing-U} with similar map density. Therefore, achieving decent performance on dense map area is still a challenging task for future map-matching research.

\vspace{-1em}

\section{Conclusion}\label{sec:conclusion}

In this paper, we conduct a comprehensive survey of the map-matching problem. We reveal the inability of all previous surveys in classifying new map-matching solutions. On top of that, we propose a new categorisation of existing methods from the technical perspective, which consists of similarity model, state-transition model, candidate-evolving model and scoring model. In addition, we list three major challenges (unnecessary detour, matching break and matching uncertainty) that the current map-matching algorithms are facing. To exemplify and demonstrate their influence on the current map-matching algorithms, we conduct extensive experiments over multiple datasets and map-matching algorithms. Overall, this paper concludes the current state of the map-matching problem and provides guidance to future research directions.

\bibliographystyle{splncs04}
\bibliography{reference}
\end{document}